\documentclass{ametsocV6.1}

\title{Remote Influences of Land Surface Temperature and their Implications for Sea Surface Temperature Patterns}

\authors{
Bosong Zhang\aff{a,b}\correspondingauthor{Bosong Zhang, bosongzhang@gmail.com} and Timothy M. Merlis\aff{a}}

\affiliation{\aff{a}{Program in Atmospheric and Oceanic Sciences,
Princeton University, Princeton, New Jersey}\\
\aff{b}{Rosenstiel School of Marine, Atmospheric, and Earth Science,
University of Miami, Miami, Florida}}
\abstract{The spatial pattern of sea surface temperature (SST) plays a central role in shaping the climate system, yet the influence of land surface temperature (LST) on large-scale SST patterns remains understudied. Using a state-of-the-art coupled climate model, we investigate the climate response to regional abrupt LST warming and time-dependent ramped-warming. We find that regional LST warming can substantially alter the mean-state SST pattern, with the response depending strongly on the location of land warming. In particular, warming over South America strengthens the zonal SST gradient across the tropical Pacific, producing a more La Niña-like mean state. Enhanced land warming increases the zonal contrast in diabatic heating, which excites stationary Rossby wave responses that strengthen alongshore winds and coastal upwelling in the eastern tropical Pacific, providing a dynamical pathway linking regional land warming to changes in the equatorial Pacific mean state. Similar remote SST responses occur for warming over North America, which is accompanied by cooling in the North Pacific, and over Central Africa, which is accompanied by cooling in the tropical Atlantic. In contrast, warming over the Maritime Continent and Tibetan Plateau produces little systematic change in large-scale SST patterns. These results demonstrate that regional LST changes can exert a substantial and geographically dependent influence on remote ocean temperatures. They further suggest that biases in both the mean state and warming trends of LST may contribute to biases in simulated SST patterns in historical simulations, highlighting the importance of accurately representing LST in climate models.}

\begin{document}
\nolinenumbers
\maketitle

%
%
%
\statement
Sea surface temperature (SST) patterns are a primary driver of global climate variability, yet the role of land surface temperature (LST) in shaping these patterns remains unclear. This study demonstrates that warming over specific land regions can systematically alter remote ocean temperatures through changes in atmospheric circulation. In particular, warming over South America strengthens the tropical Pacific zonal temperature gradient and promotes a more La Niña-like state through changes in large-scale winds and coastal upwelling. Similar remote ocean responses occur for land warming over North America and Central Africa, highlighting a broader influence of land–atmosphere interactions on ocean climate. These results reveal an overlooked pathway through which regional land warming can affect global climate patterns and suggest that biases in simulated LST mean state and warming trends may contribute to errors in simulated SST trends and patterns.
%
%
%

%








\section{Introduction}
The spatial pattern of surface temperature is a fundamental component of the climate system, shaping and interacting with atmospheric radiation, convection, and circulation across regional and global scales. In recent years, substantial attention has been directed toward the "pattern effect" \citep{stevens2016}, which is the dependence of radiative feedbacks on the spatial distribution of surface temperature. Many related studies have focused on how sea surface temperature (SST) patterns, particularly in the tropical Pacific, influence top-of-atmosphere (TOA) radiation and global climate sensitivity \citep[e.g.,][]{alessi2023,bloch2024,dong2019,quan2024,quan2025sea,williams2023,zhang2023GF,zhou2017,falasca2025fluctuation,wang2025diagnosing,kawaguchi2025responses,falasca2024data}. Overall, these studies demonstrate that changes in global-mean net TOA radiation depend on the spatial pattern of SST in atmosphere-only simulations. 

However, a persistent challenge in climate science is the inability of state-of-the-art models to reproduce observed historical SST trends, such as the cooling trend in the eastern equatorial Pacific and Southern Ocean \citep[e.g.,][]{seager2019strengthening,watanabe2021enhanced,wills2022systematic,simpson2025confronting}. Previous studies have sought to attribute spatial patterns of tropical Pacific SST trends to local physical processes, such as the upwelling of pre-industrial waters \citep{Clement96} or the temperature dependence of evaporative changes \citep{knutson95}. These local processes can be quantified via changes in surface energy budget in model simulations of future climates \citep{zhang14b} or idealized perturbation calculations motivated by trends in recent decades \citep{merlis2025perturbing}. Beyond local processes, there is interest in relating tropical Pacific SST changes to other regions. For example, there is the potential for remote responses to Southern Ocean freshwater fluxes \citep{dong2022antarctic}, Antarctic ozone changes \citep{hartmann2022antarctic}, and decadal trends in the Antarctic Circumpolar Current \citep{kang2026km} to affect the tropical Pacific. However, among all potential factors, the role of land surface temperature (LST) has remained comparatively under-scrutinized. Recent observational evidence suggests that LST may dominate global negative radiative feedbacks for monthly timescale variability \citep{thompson2025observational}, highlighting a critical need to understand land-surface contributions to climate variability. Additionally, coupled simulations indicate that increased surface energy fluxes or carbon dioxide concentrations over land can induce remote cooling in the eastern tropical Pacific \citep{zhou17b,gunther2026heating}. 

In this study, we investigate the remote influence of regional land warming on the climate system, including its effects on the spatial patterns of SST. We perform a series of idealized numerical experiments using CM4X, one of the latest fully coupled ocean-atmosphere models from NOAA’s Geophysical Fluid Dynamics Laboratory (GFDL). Our findings reveal a significant land-to-ocean teleconnection, suggesting that ``what happens over land does not stay over land''. We find that warming over South America, in particular, triggers a robust La Niña-like SST response in the Pacific with enhanced zonal SST gradient and subtropical anticyclones. Similar west-east SST-LST couplets are identified for North America and Central Africa, linked to stationary Rossby wave responses and shifts in diabatic heating. Our findings underscore the importance of representing land–atmosphere–ocean interactions for accurately simulating and predicting the evolving global climate.

\section{Methods}
\subsection{Perturbing Land Surface Temperature}

One application of atmosphere-only models is to investigate the atmospheric response to prescribed, uniform changes in SST \citep[e.g.,][]{cess90}. Another is to quantify the effective radiative forcing (ERF) of greenhouse gases, such as carbon dioxide (CO$_2$) \citep[e.g.,][]{pincus2016radiative,ramaswamy2019radiative,kramer2026radiative}. Both applications typically allow LST to respond freely, an acceptable but partly unintended feature of the experimental design. Recent studies have instead prescribed LST alongside SST when estimating CO$_2$ ERF using atmosphere-only models \citep{ackerley2016atmosphere,ackerley2018ensemble,andrews2021effective,zhang2026decoupling}. These studies show that allowing LST to evolve freely underestimates CO$_2$ ERF relative to estimates from simulations with both LST and SST prescribed. This finding highlights LST changes as an important component of the climate response and motivates further investigation of their role in the fully coupled climate system.

In this study, we examine how local LST warming affects the climate system. We select six regions to perturb LST. The selection is motivated in part by previous idealized studies. For example, \citet{gunther2026heating} examined CO$_2$-induced land warming over the entire Northern Hemisphere, all tropical land, and South America. Here, we instead consider additional localized regions, including the Maritime Continent, North America, and two South American domains: a tropical domain and a larger domain extending into the subtropics and mid-latitudes. We also include the Tibetan Plateau because of its well-established influence on the global climate system \citep[e.g.,][]{tang2025tropical}, allowing us to examine the climate response to warming over this distinctive high-elevation region (see details in Table \ref{tab:pi-experiments} for the six regions). 

We consider two types of idealized LST-warming simulations. First, we impose a uniform, abrupt 4 K warming over each selected region to evaluate the model’s response in its mean state. Second, we impose a gradual 4 K warming over South America over a 10-year period to assess the model’s long-term SST response to the imposed LST trend. In atmosphere-only simulations with prescribed SST, monthly SST input is commonly used because the ocean exhibits relatively weak high-frequency variability. In contrast, LST exhibits substantial short-term variability, including a pronounced diurnal cycle driven by solar radiation. We therefore retain the diurnal cycle of LST by prescribing surface temperature at 3-hourly intervals. 

 
\subsection{Datasets}
In this study, we use two observational datasets for land surface temperature. The first is the Berkeley Earth Surface Temperature (BEST) dataset \citep{rohde2020berkeley}, which provides globally gridded near-surface air temperature fields of 1° $\times$ 1°. In addition, we use the latest version of CRU TS \citep{harris2020version}, a widely used 0.5° $\times$ 0.5° gridded climate dataset constructed from interpolated monthly observations from weather stations worldwide (excluding Antarctica). We also use monthly data from the ERA5 reanalysis \citep{hersbach2020era5} for comparisons of model outputs. Extended Reconstructed Sea Surface Temperature (ERSST) version 6 \citep{huang2025extended} is used to diagnose the observed SST trends.

\subsection{Model and Experiments}
We use the GFDL-CM4X ocean–land-atmosphere coupled model \citep{griffies2025gfdl} to conduct our experiments. CM4X employs the $\approx 50$-km (C192) horizontal resolution configuration of AM4 \citep{zhao2020simulations} with 33 vertical levels, providing finer atmospheric grid spacing than the default AM4 setup \citep{zhao2018gfdlam4part1,zhao2018gfdlam4part2}. For the ocean, CM4X incorporates an updated MOM6 (Modular Ocean Model version 6) physics package relative to CM4.0 \citep{held2019structure}. Two versions of CM4X are documented in \citet{griffies2025gfdl} with varying horizontal resolution of MOM6: one uses a horizontal grid spacing of 0.25° (referred to as CM4X‐p25) and the other that uses a 0.125° grid (referred to as CM4X‐p125). All simulations in this work use the CM4X-p25 configuration.

We consider the following experiments:
\begin{enumerate}
\item \textbf{piControl and regional LST perturbations:} The pre-industrial control simulation is conducted with radiative forcing fixed at its year-1850 levels. The piControl run branches at model year 850 and continues through year 1200 \citep{chen2026assessing}. However, the 350-year piControl simulation used in \citet{chen2026assessing} does not output high-frequency LST. We therefore re-run the piControl simulation for 20 years, outputting 3-hourly LST to provide the baseline used in the perturbation experiments. For the regional LST perturbation simulations, an individual target region is perturbed (Table~\ref{tab:pi-experiments}), and outside the target region, all variables evolve freely. To minimize the influence of initial transient adjustment, all analyses of the +4 K perturbation experiments use only the final 15 years of each 20-year simulation. In addition to the abrupt $+4$ K LST perturbations, we performed a linear-in-time South America experiment.
\item \textbf{Historical:} In this study, we use the historical experiments described in \citet{chen2026assessing}, which include 10 ensemble members of historical simulations.
\end{enumerate}

\begin{table}
\caption{CM4X experiments used in this study. For the piControl-style experiments with fixed external forcing, we conduct both abrupt and linearly ramped LST-warming experiments. The abrupt-warming experiments are integrated for 20 years over six selected regions, while the linearly ramped warming experiment was integrated for 10 years for the larger South American domain. We also use 10 ensemble members of historical simulations from \citet{chen2026assessing}.}
\label{tab:pi-experiments}
\centering
\small
\begin{tabular}{l p{8cm} c c}
\topline
Name & Region & Members & Length \\
\midline
pi-Control & --- & 1 & 20 yr \\
CAfrica & Central Africa [0$^\circ$--60$^\circ$E, 15$^\circ$S--15$^\circ$N] & 1 & 20 yr \\
MC & Maritime Continent [90$^\circ$E--150$^\circ$E, 10$^\circ$S--10$^\circ$N] & 1 & 20 yr \\
TP & Tibetan Plateau [70$^\circ$E--110$^\circ$E, 25$^\circ$N--43$^\circ$N] & 1 & 20 yr \\
NAmerica & North America [130$^\circ$W--60$^\circ$W, 20$^\circ$N--50$^\circ$N] & 1 & 20 yr \\
SAmerica tropics & South America (tropics) [90$^\circ$W--30$^\circ$W, 10$^\circ$S--10$^\circ$N] & 1 & 20 yr \\
SAmerica large & South America [90$^\circ$W--30$^\circ$W, 40$^\circ$S--10$^\circ$N] & 1 & 20 yr \\
Linear SAmerica large & South America [90$^\circ$W--30$^\circ$W, 40$^\circ$S--10$^\circ$N] & 1 & 10 yr \\
Historical & --- & 10 & 1850--2014 \\

\botline
\end{tabular}
\end{table}

\section{Results}
\subsection{Abrupt Land Warming}
\begin{figure}
    \centering
    \includegraphics[width=\linewidth,height=0.8\textheight,keepaspectratio]{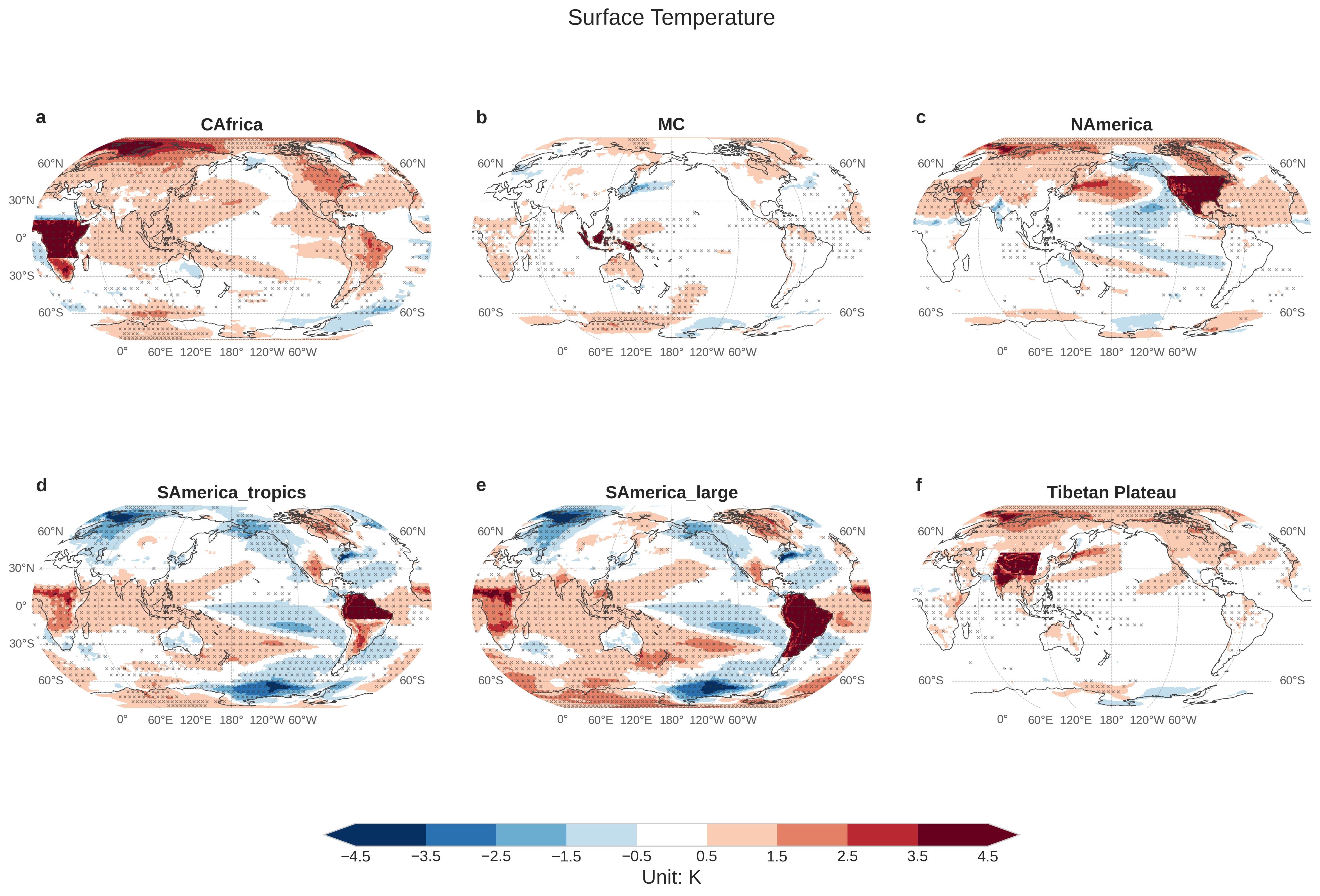}
    \caption{Annual-mean surface skin temperature response for each land-warming experiment (units: K). Crosses indicate statistical significance at the 99 confidence level based on a Welch's t-test.}
    \label{fig:ts}
\end{figure}

Figure \ref{fig:ts} shows the resulting surface temperature response. Abrupt LST warming over Central Africa has broader impacts, including remote warming over land regions on other continents. It also induces anomalous SST cooling over the southeastern Atlantic, particularly along the west coast of Africa (Figure \ref{fig:ts}a). Similar west–east pairings of SST cooling plus LST warming are found in other experiments, including land warming over South America accompanied by SST cooling over the southeast Pacific (Figure \ref{fig:ts}d \& e), and warming over North America with SST cooling over the northeastern Pacific (Figure \ref{fig:ts}c). In contrast, LST warming over the Maritime Continent (Figure \ref{fig:ts}b) and the Tibetan Plateau (Figure \ref{fig:ts}f) does not produce pronounced changes in the overall SST patterns, particularly in the zonal SST gradient.

\begin{figure}
    \centering
    \includegraphics[width=\linewidth,height=0.8\textheight,keepaspectratio]{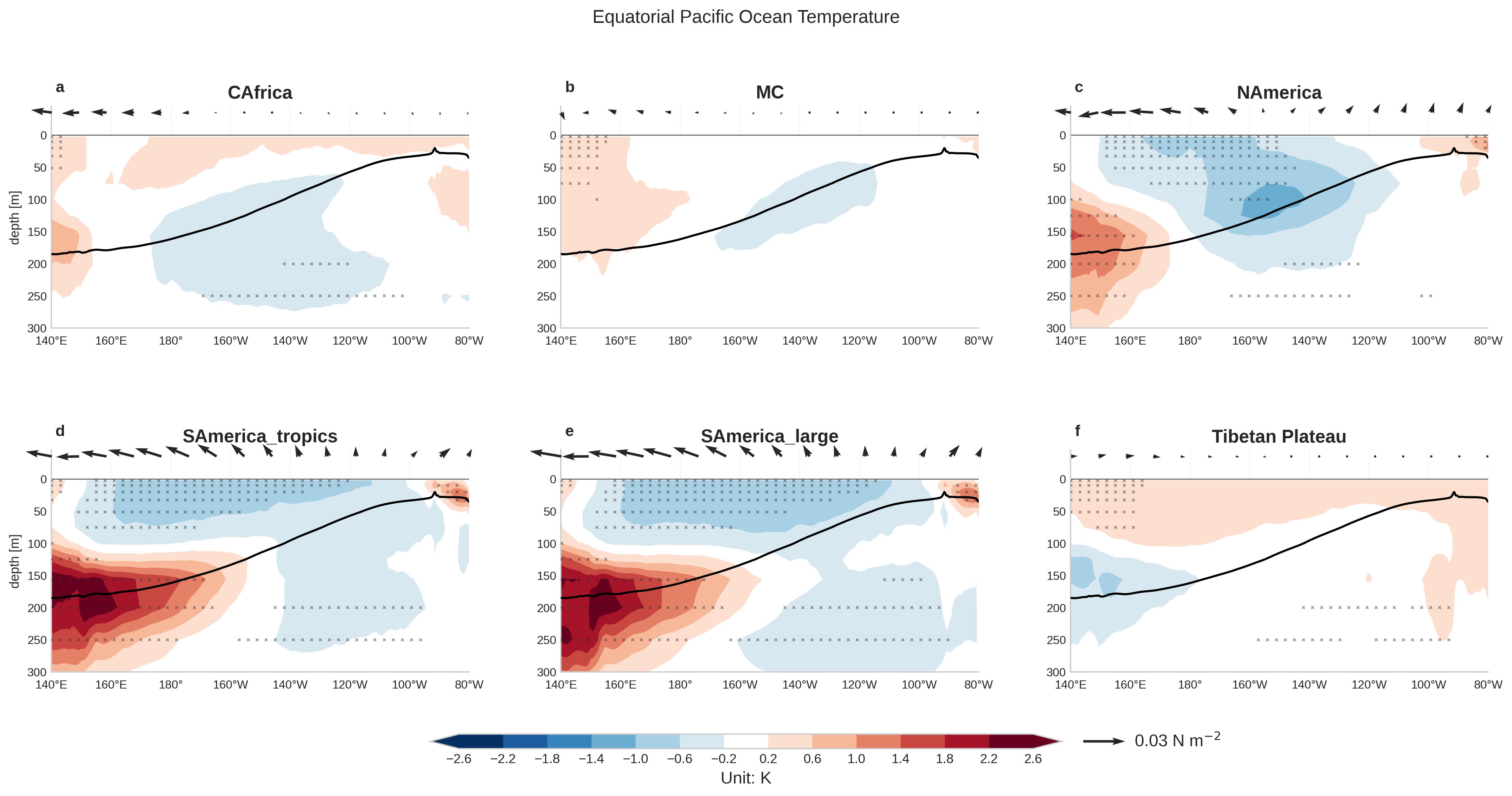}
    \caption{Equatorial Pacific Ocean temperature response to regional land warming. Depth–longitude sections show the annual-mean ocean temperature response (shading; K), averaged over 2.5°S–2.5°N, from 140°E to 80°W and over the upper 300 m. Crosses indicate statistical significance at the 99 confidence level based on a Welch's t-test. The black curve in each panel denotes the climatological 20 degree Celsius isotherm in the control simulation, providing an approximation of thermocline depth. Arrows above each panel indicate the response of equatorial surface wind stress (in $\mathrm{N \ m^{-2}}$ with the reference arrow magnitude of 0.03 $\mathrm{N \ m^{-2}}$). The horizontal and vertical components represent zonal and meridional wind stress, respectively.}
    \label{fig:ocean_t}
\end{figure}

In addition to the surface temperature response, another important aspect is vertical temperature structure below the ocean surface. Enhanced subsurface warming at depths of 100–200 m in the western Pacific, together with intensified easterly trade winds, are key indicators of the tropical Pacific warming pattern observed over the past several decades \citep[e.g.,][]{jiang24,xie2026major}. Here, we examine changes in equatorial Pacific ocean temperature within the upper 300 m. Land warming over South America is associated with enhanced subsurface warming between approximately 100 and 300 m in the Western Pacific (from 140°E to 180°), accompanied by weak cooling in the upper ocean. In addition, intensified easterly trade winds are evident in both simulations (Figure \ref{fig:ocean_t} d \& e). These changes in upper-ocean temperature and surface winds provide a distinct fingerprint of the remote influence of land warming on the tropical Pacific upper ocean, which is consistent with the surface patterns shown in Figure \ref{fig:ts}.

\begin{figure}
    \centering
    \includegraphics[width=\linewidth,height=0.8\textheight,keepaspectratio]{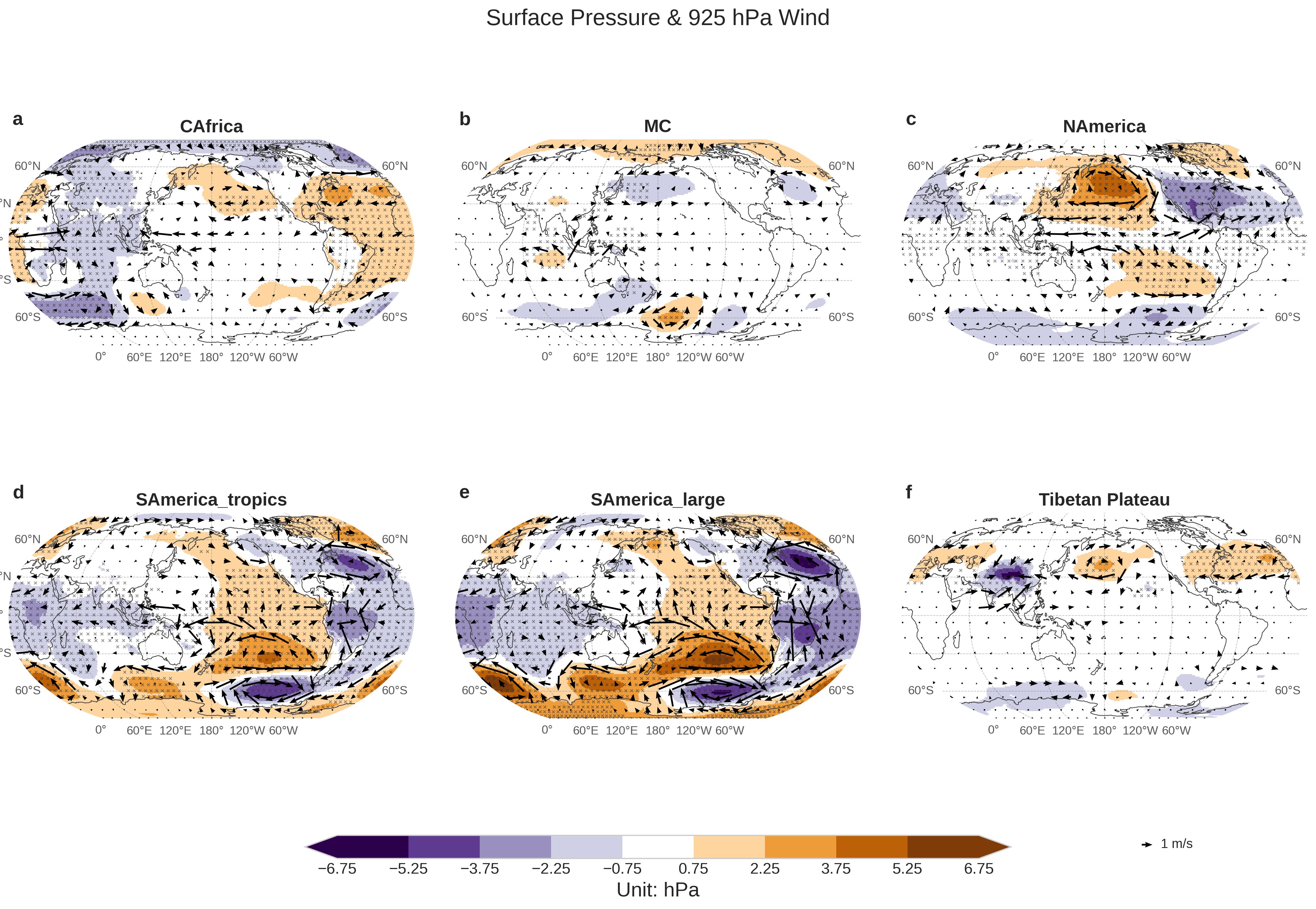}
    \caption{Same as Figure \ref{fig:ts} except the shading shows the annual-mean sea level pressure response (units: hPa) and the vectors show the 925-hPa wind response.}
    \label{fig:psl}
\end{figure}

Returning to the surface and atmospheric response, we  show the change in sea-level pressure and low-level circulation in Figure \ref{fig:psl}. The west–east SST–LST patterns shown in Figure \ref{fig:ts} are associated with corresponding west–east pressure anomalies. High pressure anomalies are found over the Atlantic with central African LST warming, although the low pressure anomalies over the continent are weaker (Figure \ref{fig:psl}a). For LST warming over North America, low-pressure anomalies develop over the continent, while the Aleutian Low over the North Pacific is weakened by anomalous high pressure (Figure \ref{fig:psl}c). In the case of South America LST warming, the subtropical high over the southeast Pacific strengthens and is accompanied by an anticyclonic circulation anomaly (Figure \ref{fig:psl}d \& e), consistent with proposed mechanisms for the formation and maintenance of Southern Hemisphere subtropical anticyclones \citep{miyasaka2010structure,rodwell2001subtropical,liu2004relationship}. In comparison, the large-scale pressure response is more spatially confined for LST warming over the Maritime Continent (Figure \ref{fig:psl}b) and the Tibetan Plateau (Figure \ref{fig:psl}f).

The physical mechanism linking the temperature, pressure, and circulation couplets is the zonal contrast in diabatic heating \citep{rodwell2001subtropical,miyasaka2005structure,miyasaka2010structure}. This can drive the Rodwell-Hoskins mechanism of subsidence to the west of anomalous diabatic heating via Rossby wave-induced subsidence \citep{rodwell96}.
Typically, previous studies examine the zonally asymmetric component of the latent heating \citep[e.g.,][]{schumacher2004tropical}. We show the precipitation response in what follows, but first consider the diabatic forcing of the \textit{moist} energetics (i.e., what one finds in a moist static energy budget) because this is `external' to the circulation \citep{emanuel94b}. It also allows for cloud-radiative induced energetic changes to the atmospheric column. Here, we assess this mechanism by examining the zonally asymmetric component of atmospheric diabatic heating (\( \mathrm{Q^{*}} \)), where the asterisk denotes the departure from the zonal mean. The diabatic heating, Q, is computed using the following equation:
\begin{equation}
\begin{split}
Q =\ & (SW_{down,\ \mathrm{TOA}} - SW_{up,\ \mathrm{TOA}} - OLR) \\
     & - (SW_{down,\ \mathrm{SFC}} - SW_{up,\ \mathrm{SFC}}
     + LW_{down,\ \mathrm{SFC}} - LW_{up,\ \mathrm{SFC}}) \\
     & + SH + LH,
\end{split}
\label{eqn-diabatic_forcing}
\end{equation}
where $SW_{down,\ \mathrm{TOA}} - SW_{up,\ \mathrm{TOA}} - OLR$ is the net radiation at TOA, $SW_{down,\ \mathrm{SFC}} - SW_{up,\ \mathrm{SFC}} + LW_{down,\ \mathrm{SFC}} - LW_{up,\ \mathrm{SFC}}$ is net radiation at the surface, $SH$ is the turbulent flux of sensible heat at the surface, and $LH$ is turbulent flux of latent heat flux at the surface.

\begin{figure}
    \centering
    \includegraphics[width=\linewidth,height=0.8\textheight,keepaspectratio]{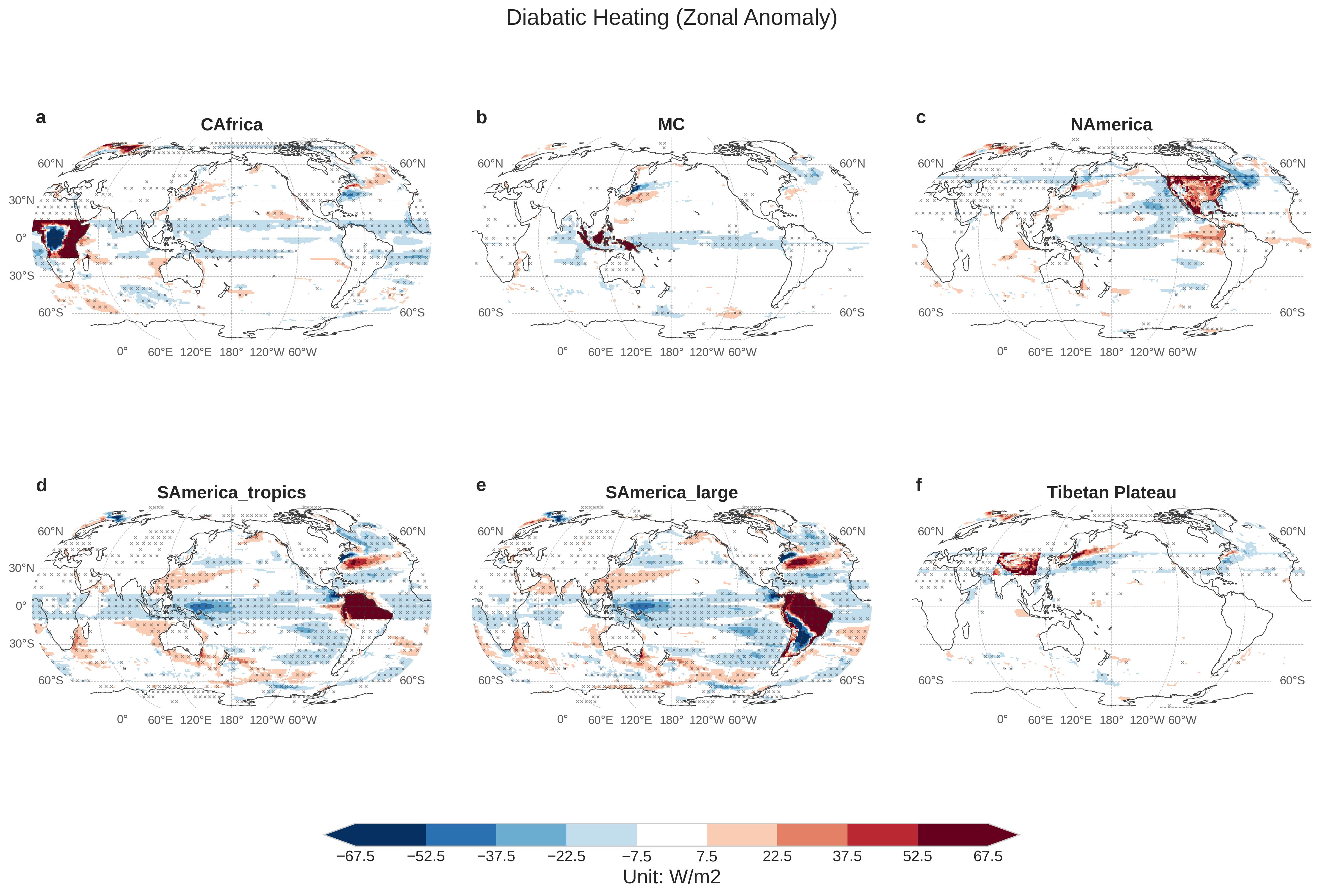}
    \caption{Same as Figure~\ref{fig:ts}, except the shading shows the response of the annual-mean zonally asymmetric component of diabatic heating (Eq.~\ref{eqn-diabatic_forcing}, units: \(\mathrm{W\,m^{-2}}\)).}
    \label{fig:q}
\end{figure}

\begin{figure}
    \centering
    \includegraphics[width=\linewidth,height=0.8\textheight,keepaspectratio]{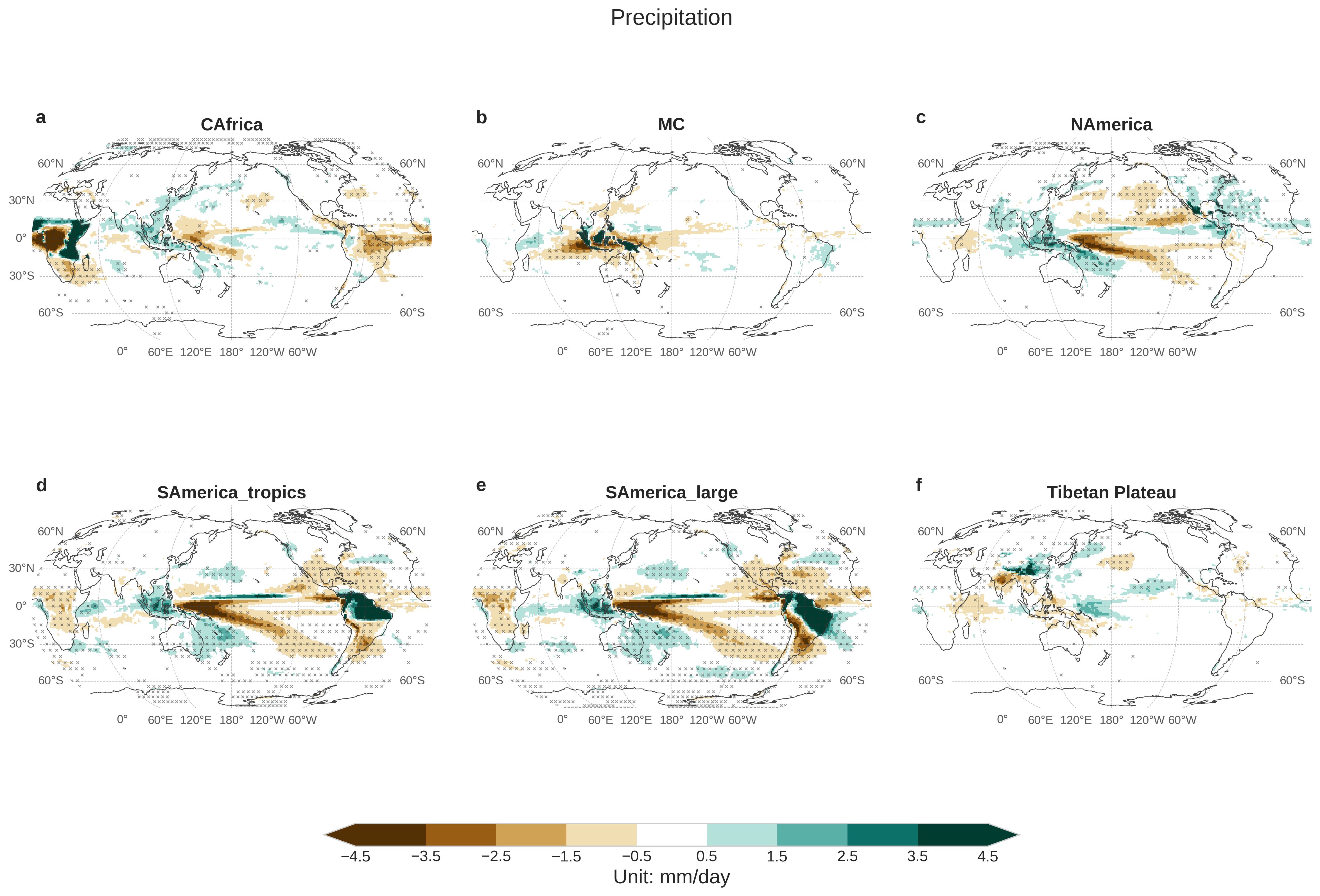}
    \caption{Same as Figure \ref{fig:ts} except the shading shows the annual-mean precipitation response (units: mm per day).}
    \label{fig:pr}
\end{figure}

As expected, \( \mathrm{Q^{*}} \) is generally positive over regions where the land surface is warmed (Figure \ref{fig:q}). However, negative anomalies are also found over land in some cases, such as Central Africa (Figure \ref{fig:q}a) and South America (Figure \ref{fig:q}e). These negative anomalies are associated with changes in precipitation (Figure \ref{fig:pr}). In the South America case, negative \( \mathrm{Q^{*}} \) appears offshore along the western coast, reflecting an enhanced zonal land–sea thermal contrast (Figure \ref{fig:q}d \& e). This contrast can excite stationary Rossby waves, strengthening subtropical highs and shifting them westward, thereby reinforcing equatorward winds along their eastern flanks \citep{rodwell96,rodwell2001subtropical,liu2004relationship}. The anomalous diabatic heating is also associated with changes in precipitation over South America. Enhanced rainfall occurs over land regions where LST is increased, while reduced rainfall is found over much of the central and eastern equatorial Pacific (Figure \ref{fig:pr}d \& e). The response of vertical velocity in the mid-troposphere has a similar pattern and relative amplitude to precipitation (not shown). This precipitation pattern is consistent with a strengthened zonal SST gradient and a more La Niña–like mean state. Stronger trade winds increase surface evaporation and enhance the advection of cooler air from higher latitudes into the subtropics. The resulting cooling in the eastern Pacific may be further amplified by the wind–evaporation–SST (WES) feedback, reinforcing cold SST anomalies in the tropical Pacific \citep[e.g.,][]{kim2022subtropical}. Similar zonal structure of \( \mathrm{Q^{*}} \) also appears for the Central Africa case (Figure \ref{fig:q}a) and the North America case (Figure \ref{fig:q}c).

Recently, \citet{gunther2026heating} identified mechanisms linking cooling in the southeast Pacific to a northward shift of the ITCZ and a westward displacement of deep convection. In CM4X, warming over the Maritime Continent enhances precipitation over land but suppresses rainfall over the surrounding ocean (Figure \ref{fig:pr}b). This spatially heterogeneous precipitation response over the Maritime Continent and adjacent oceans suggests mixed signals in the displacement of deep convection. As a result, LST warming over the Maritime Continent exerts only a weak influence on southeast Pacific SST anomalies and does not substantially strengthen the zonal SST gradient across the tropical Pacific (Figure \ref{fig:ts}b). Interestingly, we do not see an obvious equatorial Gill-model style response for the Maritime Continent perturbation. This may result from the substantial regional cancellation in precipitation and the column heating via the nearby oceanic part of the domain. In contrast, patch SST warming perturbation simulations \citep{zhang2023GF} with tropical West Pacific warming do have a response more like the Gill model (not shown). This highlights the potential complexity of the interactions that arise from LST perturbations compared to our baseline expectations. Our experimental design differs from thoat of \citet{gunther2026heating}. In their experiments, CO$_2$ concentrations were quadrupled either over the Northern Hemisphere or over a broad set of tropical land regions, including the Maritime Continent, Central Africa, and South America. In contrast, the perturbation in our experiments is confined to land areas within the Maritime Continent. Thus, the stronger teleconnections reported by \citet{gunther2026heating} may partly reflect diabatic heating over tropical South America, which is excluded from our perturbation. Differences in the model physics may also contribute to the contrasting responses.

\subsection{Gradual Land Warming}

The abrupt LST warming experiments are designed to assess the remote influence of LST anomalies on the mean state of SST. A related question is how LST forcing affects SST trends. Because South America plays an important role in modulating the Pacific zonal SST gradient, and because tropical Pacific SST patterns strongly influence global radiative feedbacks across multiple models \citep[e.g.,][]{bloch2024}, we focus here on land warming over South America. We impose a regional land warming tendency over South America for 10 years, with LST increasing linearly and reaching approximately 4 K by the end of the simulation. This experimental framework allows us to examine how ramped LST warming over South America influences the trends in SST and other climate variables. An additional advantage of the ramp-warming experiment is that it does not require a reference simulation; the response can instead be diagnosed directly from the temporal trends in the ramp-warming simulation, without defining a reference mean state.

\begin{figure}
    \centering
    \includegraphics[width=\linewidth,height=0.8\textheight,keepaspectratio]{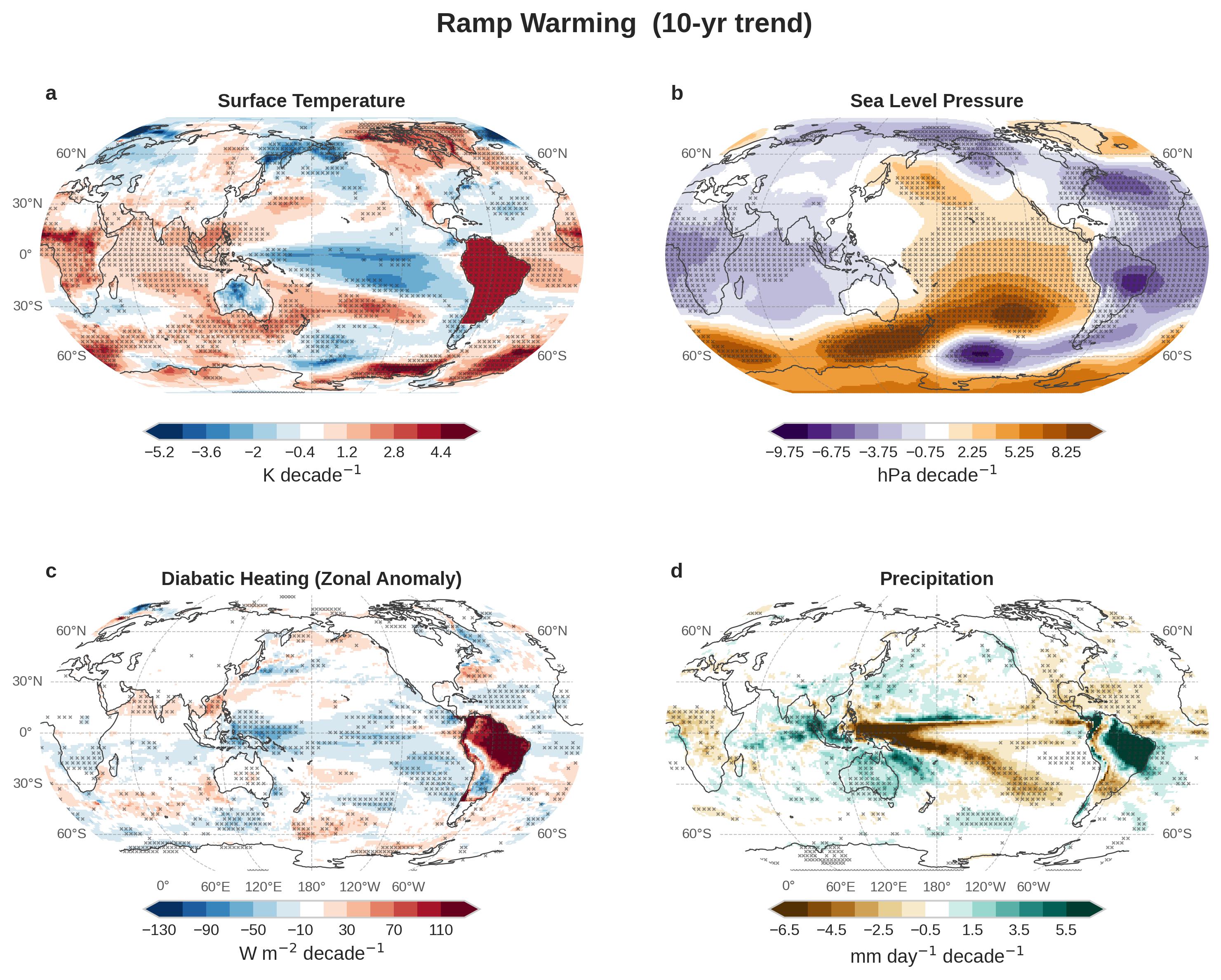}
    \caption{Maps of the linear trend in (a) surface temperature, (b) sea level pressure, (c) zonally asymmetric diabatic heating, and (d) precipitation. Crosses indicate statistical significance at the 99 \% confidence level based on a Student's t-test.}
    \label{fig:ramp_warming}
\end{figure}

Figure \ref{fig:ramp_warming} shows the trends in surface temperature, sea level pressure, \( \mathrm{Q^{*}} \) and precipitation. The SST trend closely resembles the mean SST response to abrupt land warming over South America shown in Figure \ref{fig:ts}, with pronounced cooling over the southeastern Pacific. This similarity suggests that the same land–atmosphere–ocean coupling mechanisms operate under both transient and abrupt temperature changes. These results demonstrate that warming over South America can exert a strong and persistent influence on southeastern Pacific SST, and consequently on the Pacific zonal SST gradient. The trends in the other variables shown in Figure \ref{fig:ramp_warming} also exhibit spatial patterns consistent with those found in the abrupt land-warming experiments over South America, further supporting the robustness of the physical mechanisms identified from the abrupt warming experiments.

\subsection{Implications for Historical Simulations}

\begin{figure}
    \centering
    \includegraphics[width=\linewidth,height=0.8\textheight,keepaspectratio]{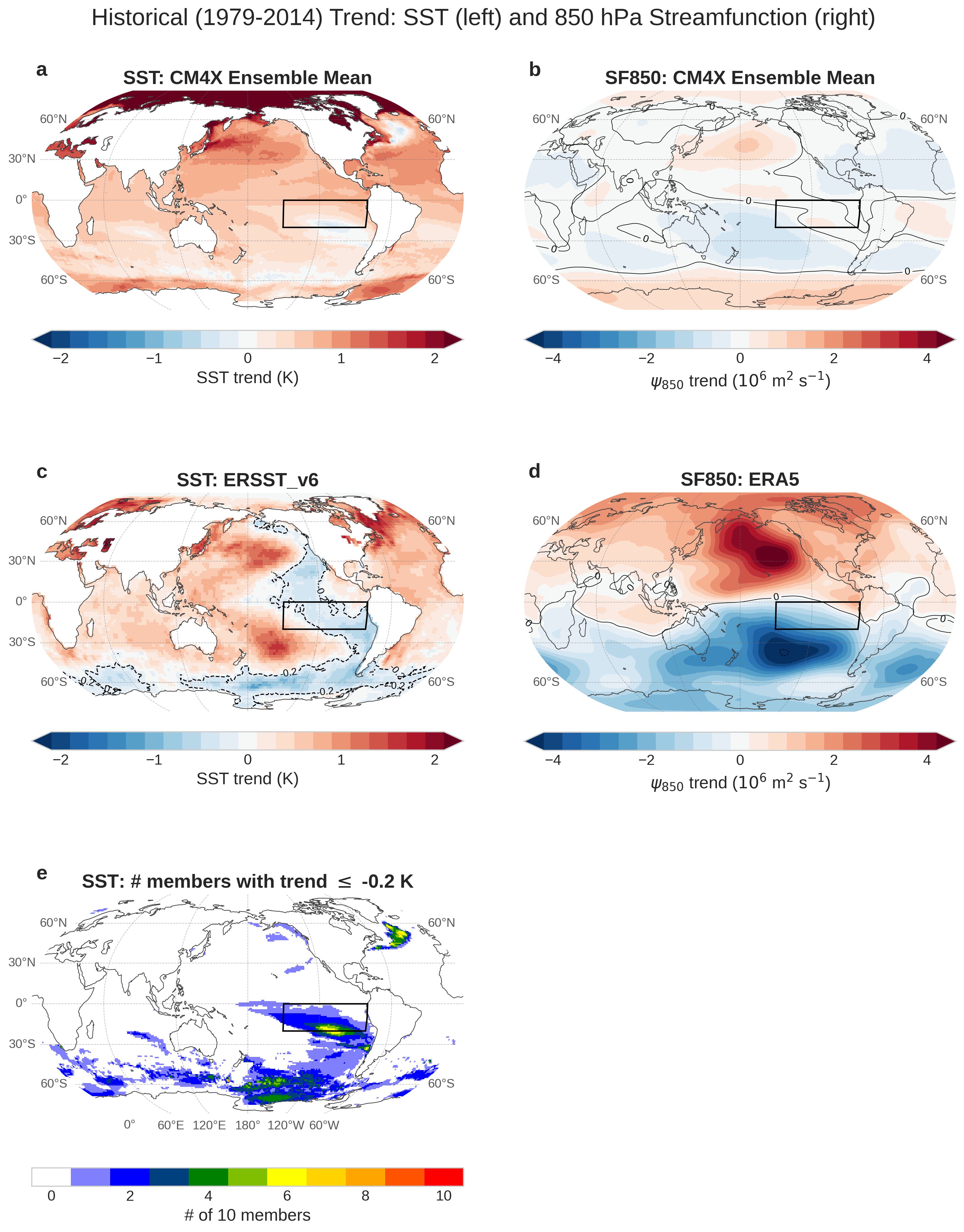}
    \caption{(a) Linear trend in SST from 1979 to 2014 based on the 10-member ensemble mean of CM4X, and (c) the corresponding trend from ERSST with regions of trends $< -0.2$ K over the period enclosed by a dashed line. (b) Linear trend in the 850-hPa streamfunction from the CM4X ensemble mean, and (d) the corresponding trend from ERA5. (e) Number of CM4X ensemble members exhibiting SST cooling greater than 0.2 K over the indicated regions.}
    \label{fig:sst_sf_trend}
\end{figure}

Previous studies have documented systematic difficulties in global climate models in reproducing observed SST and sea-level pressure trends over the historical period \citep[e.g.,][]{wills2022systematic}. The CM4X model 10-member historical simulations share similar issues. CM4X generally fails to reproduce the observed SST trends in the tropical Pacific (Figure \ref{fig:sst_sf_trend}a). Only a few of the 10 ensemble members show cooling over the southeastern tropical Pacific (Figure \ref{fig:sst_sf_trend}e). On the trend of circulation, ERA5 shows a strengthening of lower-tropospheric anticyclonic circulation in both hemispheres over the historical period (Figure \ref{fig:sst_sf_trend}d). This circulation response is largely absent from the 10-member CM4X ensemble (Figure \ref{fig:sst_sf_trend}b).

We next evaluate the model's ability to simulate LST including its mean state and trend. We compare the simulated land surface air temperature with the ERA5 reanalysis and the CRU TS and BEST observational datasets. We focus on surface air temperature because this is the variable provided by the CRU TS and BEST datasets, allowing for a consistent comparison across the three observational products.

\begin{figure}
    \centering
    \includegraphics[width=\linewidth,height=0.8\textheight,keepaspectratio]{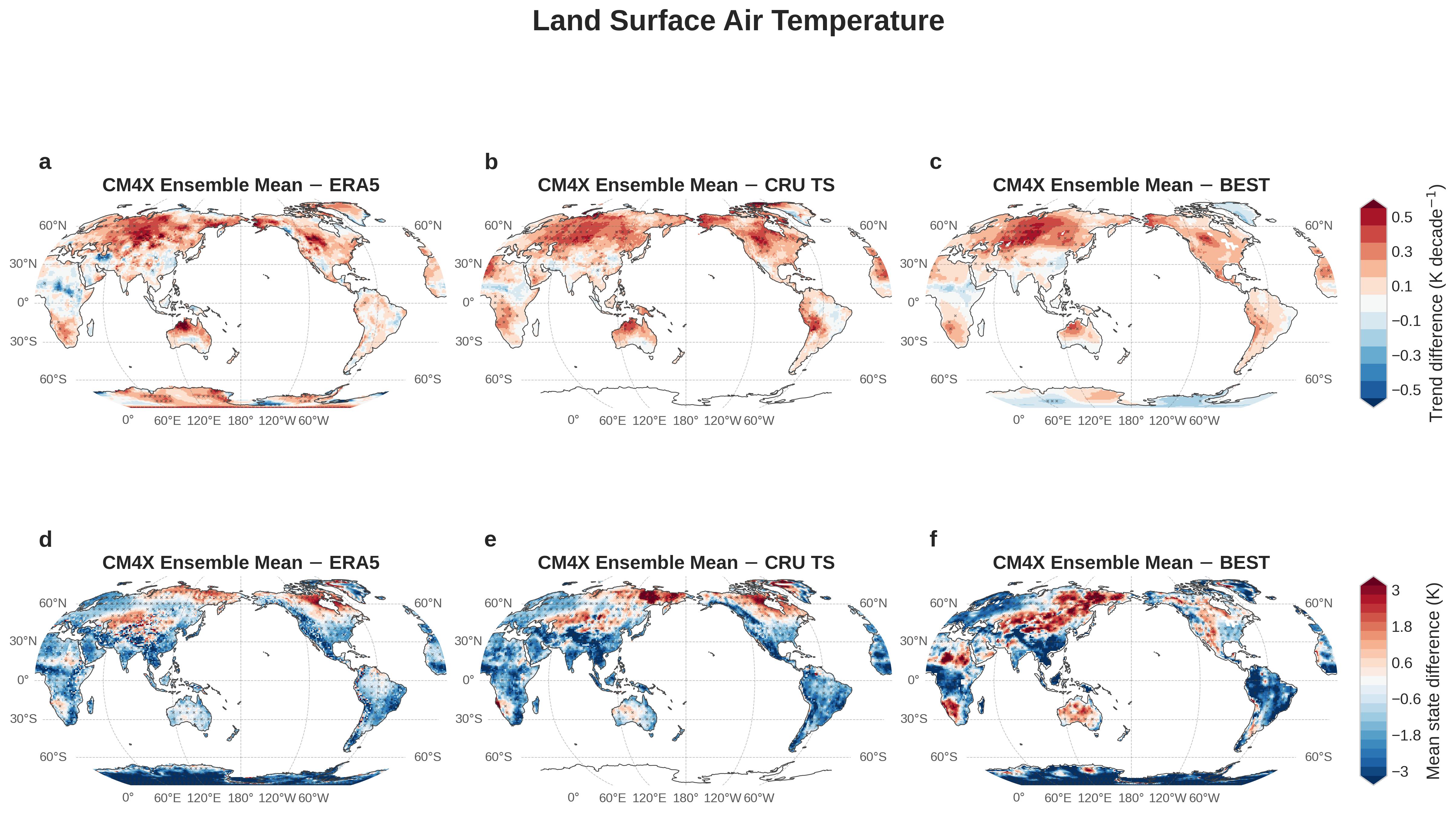}
    \caption{Linear trends in annual-mean surface air temperature over land from 1979 to 2014 for (a) the CM4X ensemble mean, (b) ERA5, (c) CRU TS, and (d) BEST. Differences between the model and ERA5, CRU TS, and BEST are shown in panels (e), (f), and (g), respectively.}
    \label{fig:land_t_trend}
\end{figure}

Over the period 1979--2014, CM4X overestimates the land warming trend compared with ERA5, CRU TS, and BEST (Figure \ref{fig:land_t_trend}a--c). However, CM4X has a cold bias in the climatological surface air temperature (Figure \ref{fig:land_t_trend}d--f). According to the findings from the idealized experiments, these two opposing biases in the trend and mean state have two implications: 1) the slightly overestimated land warming trends, particularly over South America, may contribute to producing the observed SST cooling trend over the southeastern Pacific, 2) the underestimated mean state of South America surface temperature would act against the southeast Pacific SST cooling response. Together, these opposing effects may partially offset each other.

\begin{figure}
    \centering
    \includegraphics[width=\linewidth,height=0.8\textheight,keepaspectratio]{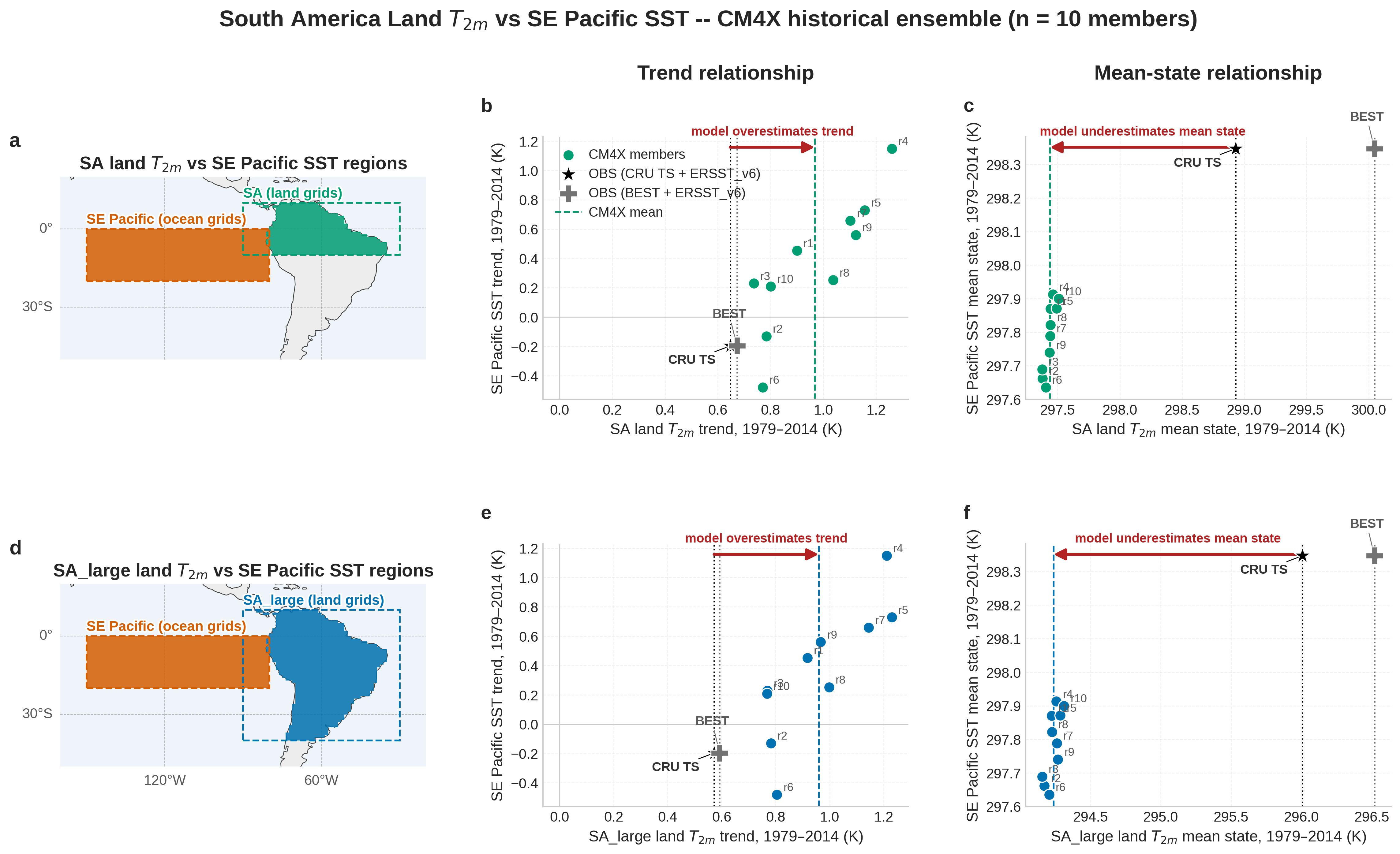}
    \caption{Relationship between domain-mean surface air temperature over land in South America and SST over the southeastern Pacific. Panel (a) illustrates the boxes defining the two regions. Panel (b) shows the relationship between their trends, while panel (c) shows the relationship between their mean states. Panels (e) and (f) show the corresponding relationships for the larger South American domains illustrated in panel (d).}
    \label{fig:scatter_trend_mean_state}
\end{figure}

Figure \ref{fig:scatter_trend_mean_state} further illustrates these two biases. When considering the tropical South American land region, all CM4X historical ensemble members overestimate the land warming trend compared to CRU TS and BEST (Figure \ref{fig:scatter_trend_mean_state}b). As in the ensemble mean, all individual CM4X members have a cold bias in the surface air temperature over the tropical South America by about 2 K (Figure \ref{fig:scatter_trend_mean_state}c). Similar results are found when considering the broader South American domain (Figure \ref{fig:scatter_trend_mean_state}e--f). We therefore argue that these two biases likely contribute in opposite directions to the overall SST trend. However, this interpretation is based on our idealized experiments. In the historical simulations, remote biases can also contribute to the SST response and therefore cannot be excluded. 

To reduce model biases in land temperature, one could in principle use observed surface temperature to bias-correct the model simulations. However, several challenges arise. First, observational datasets such as BEST and CRU TS generally provide near-surface air temperature, whereas typical land models address the surface skin temperature. These quantities can differ substantially, particularly over land, where the surface skin temperature responds directly to the surface energy balance while near-surface air temperature is additionally influenced by turbulent exchange and boundary-layer processes. Directly comparing the two therefore introduces a mismatch between the modeled and observed quantities. Second, observational datasets are typically available only at monthly or longer temporal resolutions, while LST has a pronounced large diurnal cycle. The mismatch in both the definition and temporal resolution of the temperature variable complicates the application of observationally based bias corrections to model-simulated LST. Given these challenges, we leave a more comprehensive evaluation and correction of model LST biases over the historical period as an outstanding issue that warrants further investigation.

\section{Summary and Discussion}

What happens over land does not stay over land. Our experiments with the GFDL CM4X model demonstrate that regional changes in land surface temperature (LST) can exert a substantial influence on the broader climate system, including the spatial pattern of sea surface temperature (SST). We find that LST warming over South America strengthens the tropical Pacific zonal SST gradient and produces a more La Niña-like mean state, with pronounced cooling over the southeastern Pacific. Similar remote responses are found for LST warming over North America, which is accompanied by cooling over the North Pacific, and for warming over Central Africa, which is associated with cooling over the tropical southeastern Atlantic. These responses are mediated by changes in diabatic heating and the associated Rossby wave response, which modifies large-scale atmospheric circulation and influences surface winds and ocean upwelling. In contrast, LST warming over the Maritime Continent and Tibetan Plateau produces comparatively weak SST responses, underscoring the importance of the geographic location and atmospheric teleconnection pathways through which land warming affects the ocean.

Several caveats should be considered when interpreting these results. First, the magnitude and spatial structure of the responses may be model-dependent, reflecting differences in the representation of convection, land–atmosphere coupling, atmospheric circulation, and ocean dynamics across climate models. Second, our idealized experimental design prescribes regional LST warming without explicitly introducing an external forcing mechanism, such as a regional increase in atmospheric CO$_2$ concentration as in \citet{gunther2026heating}. This design allows us to isolate the climatic response to LST changes, but it does not capture the full set of processes through which external forcing generates regional land warming. In the real climate system, changes in radiation, soil moisture, vegetation, and atmospheric composition can simultaneously influence LST and atmospheric circulation. Additional modeling and observational evidence is therefore needed to assess the robustness and real-world relevance of the mechanisms identified here.

Nevertheless, these idealized experiments provide new insight into how regional land temperature changes can influence the climate system well beyond the land surface itself. Land surface temperature patterns, which have received considerably less attention than SST patterns, can exert a substantial and persistent influence on remote ocean temperatures in coupled climate simulations. The associated stationary Rossby wave responses highlight the important role of land–atmosphere–ocean interactions in shaping both the mean climate state and its spatial trends. In particular, our results suggest that biases in simulated LST may contribute to biases in SST patterns and trends, providing a potential pathway for understanding some of the regional SST biases in historical climate simulations. Future multi-model experiments, together with observationally constrained analyses, will be important for determining the robustness of these responses and quantifying the contribution of LST-driven mechanisms to historical SST variability and trends. More broadly, our findings suggest that accurately representing the spatial pattern and evolution of LST may be important not only for simulating continental climate, but also for improving our understanding and simulation of remote ocean and global climate responses.

%

%

\clearpage
\acknowledgments

We thank Yan-Ting Chen, Tra Dinh, Niki Zadeh, and Raphael Dussin for contributions for the CM4X setup. We acknowledge support from the National Oceanic and Atmospheric Administration, U.S. Department of Commerce under award NA23OAR4320198. The statements, findings, conclusions, and recommendations are those of the author and do not necessarily reflect the views of the National Oceanic and Atmospheric Administration, or the U.S. Department of Commerce.
%
%
\datastatement
The simulations presented here were performed using High Performance Computing resources provided by the Cooperative Institute for Modeling the Earth System, with help from the Princeton Institute for Computational Science and Engineering. BEST data can be found at https://berkeleyearth.org/data/. CRU TS can be found at https://crudata.uea.ac.uk/cru/data/hrg/. The code and a subset of the model data generated in this study are available at https://zenodo.org/uploads/23398361.

%






%



\bibliographystyle{ametsocV6}
\bibliography{references}

\end{document}